# 4MOST – 4-metre Multi-Object Spectroscopic Telescope

Roelof S. de Jong[a], Olga Bellido-Tirado[a], Cristina Chiappini[a], Éric Depagne[a], Roger Haynes[a,n], Diane Johl[a], Olivier Schnurr[a], Axel Schwope[a], Jakob Walcher[a], Frank Dionies[a], Dionne Haynes[a,n], Andreas Kelz[a], Francisco S. Kitaura[a], Georg Lamer[a], Ivan Minchev[a], Volker Müller[a], Sebastián E. Nuza[a], Jean-Christophe Olaya[a,n], Tilmann Piffl[a], Emil Popow[a], Matthias Steinmetz[a], Uğur Ural[a], Mary Williams[a], Roland Winkler[a], Lutz Wisotzki[a], Wolfgang R. Ansorge[b], Manda Banerji[c], Eduardo Gonzalez Solares[c], Mike Irwin[c], Robert C. Kennicutt, Jr.[c], David King[c], Richard McMahon[c], Sergey Koposov[c], Ian R. Parry[c], Nicholas A. Walton[c], Gert Finger[d], Olaf Iwert[d], Mirko Krumpe[d], Jean-Louis Lizon[d], Mainieri Vincenzo[d], Jean-Philippe Amans[e], Piercarlo Bonifacio[e], Mathieu Cohen[e], Patrick Francois[e], Pascal Jagourel[e], Shan B. Mignot[e], Frédéric Royer[e], Paola Sartoretti[e], Ralf Bender[f], Frank Grupp[f], Hans-Joachim Hess[f], Florian Lang-Bardl[f], Bernard Muschielok[f], Hans Böhringer[g], Thomas Boller[g], Angela Bongiorno[g], Marcella Brusa[g], Tom Dwelly[g], Andrea Merloni[g], Kirpal Nandra[g], Mara Salvato[g], Johannes H. Pragt[h], Ramón Navarro[h], Gerrit Gerlofsma[h], Ronald Roelfsema[h], Gavin B. Dalton[i,o], Kevin F. Middleton[i], Ian A. Tosh[i], Corrado Boeche[j], Elisabetta Caffau[j], Norbert Christlieb[j], Eva K. Grebel[j], Camilla Hansen[j], Andreas Koch[j], Hans-G. Ludwig[j], Andreas Quirrenbach[j], Luca Sbordone[j], Walter Seifert[j], Guido Thimm[j], Trifon Trifonov[j], Amina Helmi[k], Scott C. Trager[k], Sofia Feltzing[l], Andreas Korn[m], Wilfried Boland[n]

[a]Leibniz-Institut für Astrophysik Potsdam (AIP), An der Sternwarte 16, D-14482 Potsdam, Germany, [b]RAMS-CON Management Consultants, Assling, Germany, [c]University of Cambridge, United Kingdom, [d] European Southern Observatory, Garching bei München, Germany, [e] GEPI, Observatoire de Paris-Meudon, CNRS, Univ. Paris Diderot, France, [f] Universität-Sternwarte München, Germany, [g] Max-Planck-Institut für extraterrestrische Physik, München, Germany, [h]NOVA-ASTRON, Dwingeloo, the Netherlands, [i] Rutherford Appleton Lab., United Kingdom, [j]Zentrum für Astronomie der Universität Heidelberg, Germany, [k] Kapteyn Astronomical Institute, Groningen, the Netherlands, [l] University of Lund, Sweden, [m] University of Uppsala, Sweden, [n]innoFSPEC, Potsdam, Germany, [o]University of Oxford, United Kingdom, [n]NOVA, the Netherlands

**ABSTRACT**

The 4MOST consortium is currently halfway through a Conceptual Design study for ESO with the aim to develop a wide-field (>3 square degree, goal >5 square degree), high-multiplex (>1500 fibres, goal 3000 fibres) spectroscopic survey facility for an ESO 4m-class telescope (VISTA). 4MOST will run permanently on the telescope to perform a 5 year public survey yielding more than 20 million spectra at resolution R~5000 ($\lambda$=390–1000 nm) and more than 2 million spectra at R~20,000 (395–456.5 nm & 587–673 nm). The 4MOST design is especially intended to complement three key all-sky, space-based observatories of prime European interest: Gaia, eROSITA and Euclid. Initial design and performance estimates for the wide-field corrector concepts are presented. Two fibre positioner concepts are being considered for 4MOST. The first one is a Phi-Theta system similar to ones used on existing and planned facilities. The second one is a new R-Theta concept with large patrol area. Both positioner concepts effectively address the issues of fibre focus and pupil pointing. The 4MOST spectrographs are fixed configuration two-arm spectrographs, with dedicated spectrographs for the high- and low-resolution fibres. A full facility simulator is being developed to guide trade-off decisions regarding the optimal field-of-view, number of fibres needed, and the relative fraction of high-to-low resolution fibres. The simulator takes mock catalogues with template spectra from Design Reference Surveys as starting point, calculates the output spectra based on a throughput simulator, assigns targets to fibres based on the capabilities of the fibre positioner designs, and calculates the required survey time by tiling the fields on the sky. The 4MOST consortium aims to deliver the full 4MOST facility by the end of 2018 and start delivering high-level data products for both consortium and ESO community targets a year later with yearly increments.

**Keywords:** Wide-field multi-object spectrograph facility, VISTA telescope, NTT telescope, fibre postioner, wide field corrector, facility simulator, Gaia, eROSITA, Euclid

# 1. INTRODUCTION

The 4MOST consortium aims to provide the ESO community with a wide-field fibre-fed spectroscopic survey facility on the VISTA telescope with a large enough Field-of-View (FoV) to survey a large fraction of the Southern sky in a few years, a multiplex and spectral resolution high enough to detect chemical and kinematic substructure in the stellar halo, bulge and disks of the Milky Way, and enough wavelength coverage to secure receding velocities of extra-galactic objects over a large range in redshift. Such an exceptional instrument enables many science programs, but our design is especially intended to complement three key space-based observatories of prime European interest, Gaia, eROSITA and Euclid. Such a facility has been identified as of critical importance in a number of recent European strategic documents[2][6][7][16] and forms the perfect complement to the many other large area "imaging" survey projects around the world (e.g., VISTA, VST, DES, SkyMapper, Pan-Starrs, LSST, SKA (and precursors MeerKAT, ASKAP, MWA), WISE, GALEX).

The 4MOST design philosophy is based on the notion that 4MOST is not just an instrument, but is a *survey facility*, meaning that:

- **4MOST runs all the time**: there will be minimal instrument changes, 4MOST will be running almost all of the time on the telescope during its main 5 year survey,

- **4MOST provides a total package**: the target selection, operations and survey strategy, instrument capabilities, and high level data product delivery are all part of facility and are optimally tuned to compliment each other,

- **One design fits many science cases**: the design and operations will minimize the constraints on science cases that need optical spectroscopy, but the number of observing modes (e.g., spectrograph configurations) should be kept to a minimum (preferably one). The goal is to deliver a general-purpose, reliable, but simple instrument design, operations concept and data analysis software that is well suited to most science cases. To increase efficiency all science cases will be running at the same time in parallel, all the time.

4MOST is currently in the Conceptual Design phase, which started September 2011 and will run till February 1, 2013. During the first half of the study comparison designs were developed for both the NTT and VISTA telescopes. Based on a trade-off comparing science performance, cost, risk, and schedule for both telescopes the consortium preferred to develop 4MOST further on VISTA, which was endorsed by ESO. The technical development institutes are the Leibniz-Institut für Astrophysik, Potsdam (AIP), Universität-Sternwarte München, Max-Planck-Institut für extraterrestrische Physik, Landessternwarte Heidelberg in Germany, the University of Cambridge and Rutherford Appleton Lab in the United Kingdom, the Observatoire de Paris à Meudon in France, ASTRON in the Netherlands, and the European Southern Observatory. Additional science support is provided by from the Uppsala and Lund Universities in Sweden and the University of Groningen in the Netherlands.

In Section 2 we describe the main science drivers of 4MOST, which drives the instruments specifications laid out in Section 3. The main features of the operations concept are described in Section 4. In Section 5 we present concepts for selected subsystems and the 4MOST facility simulator is described in Section 6. In Section 7 we conclude with the further development program.

# 2. SCIENCE DRIVERS

## 2.1 Gaia

ESA's Gaia satellite, due for launch fall 2013, will provide distances from parallaxes and space kinematics from proper motions for more than one billion Milky Way stars down to $m_V$~20 mag. Gaia will also provide radial velocities and astrophysical characterization for about 150 million stars, but its sensitivity is limited to $m_V$~12–16 mag. The Gaia limits dependent strongly on stellar spectral type, because the spectrograph only covers the CaII-triplet region at 847–874 nm.

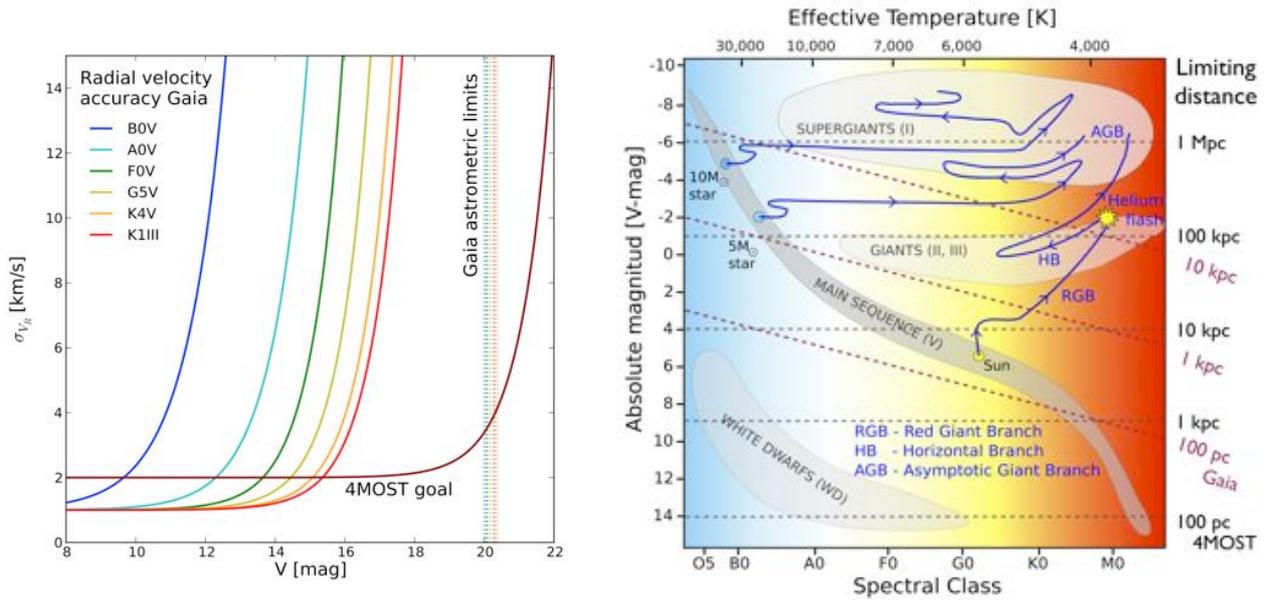

Figure 1: *Left*) The 4MOST goal radial velocity accuracy compared to the Gaia end of mission accuracy as function of stellar apparent magnitude. Our aim is to match the spectroscopic magnitude limits of 4MOST to the astrometric limits of Gaia, thereby enabling 6D-phase space studies to Gaia's limits. *Right*) Limiting distances for radial velocity measurements with Gaia (maroon diagonal) and 4MOST (black horizontal) overlaid on an HR-diagram. 4MOST can measure sun-like stars to nearly the Milky Way center, RGB stars to 100 kpc, and massive stars throughout the Local Group, substantially expanding on Gaia's spectroscopic view. Distance limits for the 4MOST high-resolution spectroscopy are about 4x smaller.

Figure 1 shows how, by covering the full optical wavelength region, the 4MOST instrument complements Gaia where it lacks spectroscopic capabilities, enabling the possibility to obtain full 6D space coordinate information and chemical characterization of objects throughout the Milky Way. Large area surveys of faint Galactic stellar objects will enable us to elucidate the formation history of the Milky Way. Models of hierarchical galaxy formation predict large amounts of dynamical substructure in the Milky Way halo that 4MOST can detect through measuring Red Giant Stars (Figure 2). Furthermore, with 4MOST we will be able to determine the three-dimensional Galactic potential and its substructure, discern the dynamical structure of the Milky Way disk and measure the influence of its bar and spiral arms, measure the Galactic assembly history through chemo-dynamical substructure and abundance pattern labelling, and find 1000s of extremely metal-poor stars to constrain early galaxy formation and the nature of the first generations of stars in the Universe.

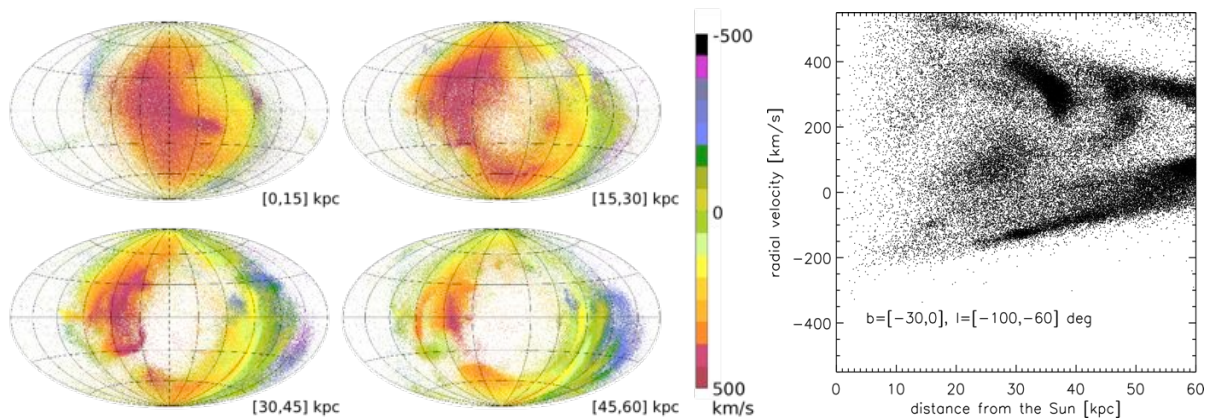

Figure 2: Spatial and radial velocity substructure distribution of red giant branch stars on the sky for a stellar halo formed in the Aquarius project cosmological simulations[5][10]. The projection corresponds to stars located in the inner halo in 4 distance bins (left) and in one direction on the sky (right) and clearly demonstrates the large amount of substructure that becomes apparent thanks to the opening of a new phase-space dimension (in this case, the line-of-sight velocity).

## 2.2 eROSITA

eROSITA (extended ROentgen Survey with an Imaging Telescope Array[14]) will perform all-sky X-ray surveys in the years 2014 to 2018 to a limiting depth a factor 30 deeper than the ROSAT all-sky survey with broader energy coverage, better spectral resolution, and better spatial resolution. 4MOST will be used to survey the >50,000 Southern X-ray galaxy clusters that will be discovered by eROSITA, measuring 3–30 galaxies in each cluster. These galaxy cluster measurements determine the evolution of galaxy populations in clusters, yield the cluster mass evolution, and provide highly competitive constraints on Dark Energy evolution. 4MOST will also enable us to determine the nature of >1 million AGNs, thus constraining the cosmic evolution of active galaxies to z=5 (Figure 3). With 4MOST we will characterize several 100,000 dynamo- and accretion-powered Galactic X-ray emitters, thereby uncovering the high energy Milky Way sources and constraining evolutionary channels of stellar populations.

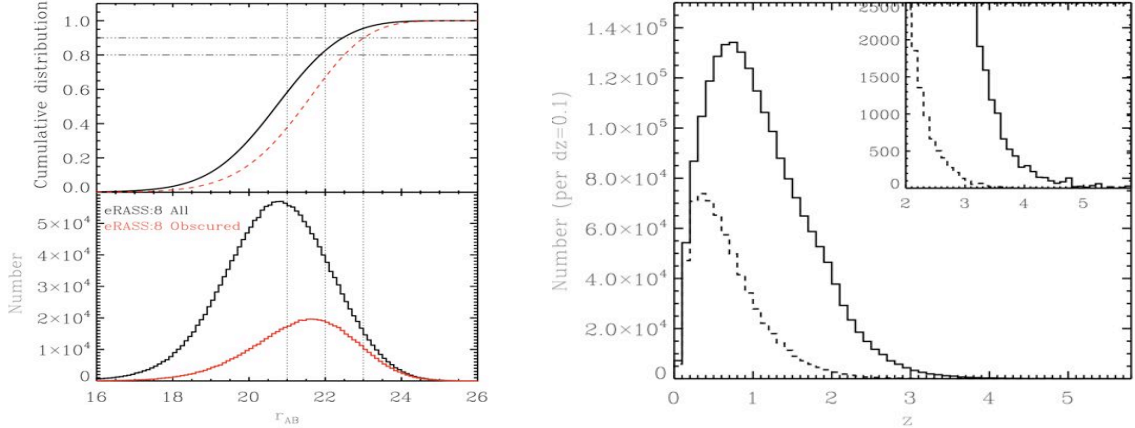

Figure 3: (*Left*) Bottom panel: expected distribution of r-band optical magnitudes for the counterparts of eROSITA AGN detected at the end of the 4-year all-sky survey. In black we show the entire 0.5-2 keV selected sample, in red only the obscured AGN. Top panel: Cumulative distribution. Vertical lines at $r_{AB}$= 21, 22, 23 and horizontal ones at 0.8 and 0.9 help guiding the eye. Even considering only obscured eROSITA AGN, reaching down to $r_{AB}$~22.5 would ensure a very high completeness level of about 80%. (*Right*) Expected redshift distribution of X-ray detected AGN in the 4-years eROSITA all-sky survey (0.5-2 keV selected). Solid lines are for all sources, dashed for those with optical spectra with narrow emission lines and/or of elliptical galaxy type. The inset shows a zoom of the high-redshift population.

## 2.3 Other science cases

Many other science cases will be enabled by 4MOST but are currently not driving extra design constraints. Examples in this non-exhaustive list include: obtaining the sample of >100,000 galaxy redshifts needed to calibrate the Euclid photometric redshifts, determine cosmological constraints from the Large Scale Structure of galaxies (e.g. baryon Acoustic oscillations), measuring galaxy evolution from redshift surveys to z~1.5, investigate the nature of radio galaxies found by the SKA telescope, follow-up of LSST and Euclid transients, determine the star formation history of the Milky Way from spectroscopy of all 100,000 (Southern) white dwarfs to be discovered by Gaia, characterize the chemo-dynamics of the Magellanic Clouds by obtaining spectra of 1 million Giant stars, elucidate the nature of peculiar variable stars discovered by Gaia and LSST, and constrain the ages of stars using astro-seismology from e.g. Corot and Kepler.

## 3. INSTRUMENT SPECIFICATION

To achieve the main science goals described above, 4MOST shall be able to obtain:

- radial velocities of ≤2 km/s accuracy of the faintest stars observed by Gaia by obtaining spectra of 19.5 r-mag stars with S/N=10 per Ångström at resolution R=5000,

- abundances of up to 15 chemical elements of 16 V-mag stars by obtaining spectra with S/N=140 per Ångström at resolution R=20000,

- redshifts of 22 r-mag galaxies and AGN.

The science requires that, within a 5 year survey, 20 (goal 30) million targets shall be observed at R~5000 and 2 (goal 3) million objects at R~20,000. In this period at least a 16,000 (goal 20,000) deg$^2$ area on the sky shall be covered at least two times (goal three times). Based on these User Requirements the main Instrument Specifications listed in Table 1 have been defined that drive the technical design.

Table 1: Main 4MOST instrument specifications.

| Specification | Requirement | Goal |
|---|---|---|
| Spectral resolution<br>  Low Resolution Spectrograph<br><br>  High Resolution Spectrograph | <br>R>5000 over full λ range<br><br>R>18,000 over full λ range,<br>R>20,000 average over full λ range | <br>R>7000 @ 800 nm,<br>R>5000 over full λ range<br>R>18,000 over full λ range,<br>R>20,000 average over full λ range |
| Wavelength coverage<br>  Low Resolution Spectrograph<br>  High Resolution Spectrograph | <br>400–900 nm<br>395–456.5 nm & 587–673 nm | <br>390–1050 nm<br>390–456.5 nm & 585–677 nm |
| Photon detecting percentage | >15% in 1.1 arcsec seeing | >20% in 1.1 arcsec seeing |
| Spectral crosstalk | <0.05% after data reduction | <0.02% after data reduction |
| Fibre aperture diameter | 1.5″±0.1″ | 1.5″±0.1″ |
| Field-of-View in hexagon | >3 deg$^2$ | >5 deg$^2$ |
| Fibre multiplex per pointing<br>  Low Resolution Spectrograph<br>  High Resolution Spectrograph | <br>>1500<br>>165 | <br>>3000<br>>330 |
| Observing efficiency | Overhead < 20% | Overhead < 10% |
| Available sky area | Zenith angle: 10°–52° | Zenith angle: 5°–52° |

## 4. OPERATIONS CONCEPT

To reach maximum impact, 4MOST will run continuously on VISTA performing a 5-year Public Survey (with no visiting astronomers) delivering ≥20 million (goal 30 million) spectra over 15,000–20,000 deg$^2$, which is an order of magnitude larger than the SDSS spectroscopic survey at >2.5× the spectral resolution. The targets selected for this Public Survey will be determined through a combination of open calls to the ESO astronomical community and the consortium GTO time, with all surveys running in parallel. Observing objects from many survey catalogues simultaneously at each pointing enables surveys that require tens of thousands objects spread sparsely over the sky. Such surveys have too few targets to use all 4MOST fibers in one pointing, but are too large to be performed in standard observing modes with existing facilities, so are optimally performed in parallel with other science programs.

During science operations the basic observational unit of 4MOST is a field. A field is a region of the sky that fits within the 4MOST field-of-view. Fields can be placed next to each other and can overlap to a small degree to allow for repeat observations. A strategy of several subexposures within one pointing, with repositioning of some of the fibres for each subexposure, is followed. The positioner therefore must be able to reposition certain fibres to new objects between two tiles within the readout time of the CCDs.

The calibration plan is an integral part of the operations concept. The general idea of the 4MOST calibration plan is to minimize calibration time during the night and to thus maximize scientifically useful time. Nevertheless, the other overarching concern is that badly calibrated data is scientifically less useful and night-time calibration will be possible for testing and validation purposes. Throughput calibration is especially critical for sky subtraction and various methods are being considered to monitor the throughput variation due to fibre bending in the positioner and telescope.

The data management system will include data quality control loops on time scale of minutes (technical faults), days (spectral quality), and half year (survey progress). The consortium will make all data, including high-level science products, available to the general public in yearly increments through a high-quality database system.

# 5. INSTRUMENT DESIGNS

The 4MOST instrument follows in general the design of previous high multiplex, fibre-fed spectrograph systems. A schematic layout of the system is shown in Figure 4. In this section we highlight the conceptual designs for the Wide Field Corrector, the Fibre Positioner and Fibre Feed, and the High- and Low-Resolution Spectrographs.

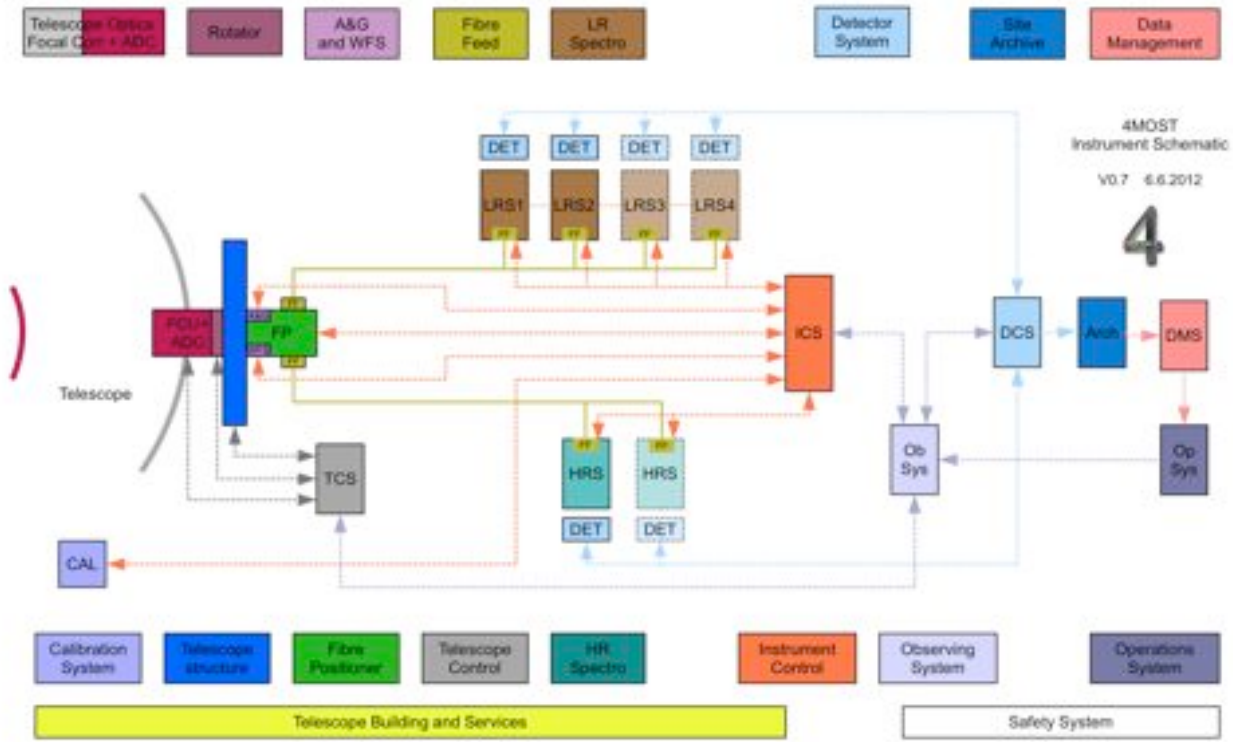

Figure 4: Schematic instrument layout with main data flows indicated. The number High- and Low-Resolution Spectrographs (HRS and LRS) is still a design trade-off.

## 5.1 Wide field correctors

Wide Field Corrector (WFC) designs were developed for both the VISTA and NTT telescopes. For the NTT an innovative "forward-Casssegrain" design was developed at the University of Munich with a focal surface above the primary mirror, as the hole in the NTT primary mirror is not large enough to host a wide field in the conventional Cassegrain position. The NTT design requires a new secondary mirror and a three-lens corrector and includes an Atmospheric Dispersion Correction (ADC) provided by a lateral movement of the middle lens in telescope elevation direction. Details of this design are presented somewhere else in these proceedings by Grupp et al.[8].

The VISTA WFC design, developed at the Institute of Astronomy (Cambridge), makes use of the current instrument interchange capability of VISTA (Figure 5) and provides a focal surface at the conventional Cassegrain focus. Designs were developed for 2.2°, 2.5°, and 3.0° diameter fields-of-view (FOV), providing about 3.0, 4.25, and 6.1 deg$^2$ FOV over the encircled hexagon focal area. In the second half of the design study the optimal FOV will be determined using the 4MOST facility simulator and cost&risk criteria but given the large science gains versus relatively small extra costs we expect the FOV will converge toward the upper range of possibilities studied so far. The optical design for the 2.5° diameter design is shown in Figure 6 (left), with image quality indicated by spot and ensquared energy diagrams in Figure 7. The coupling efficiency of light from the atmosphere into a fibre was calculated using Zemax Geometric Image Analysis tool, in essence a broadband ray-tracing tool, taking in account coatings, polarization, and bulk absorption in the material. The wavelength dependent coupling is shown in Figure 6 (right) for a range of Gaussian-shaped "seeing" functions. The design is not aberration-limited. The wavelength dependence is mostly due to the coatings (Aluminium coating taken for M1 and M2, 99% transmission coating for all glass-air and air-glass interfaces), and the losses in throughput are mainly due the seeing-limited coupling in the fibres, and to bulk-absorption in the glass.

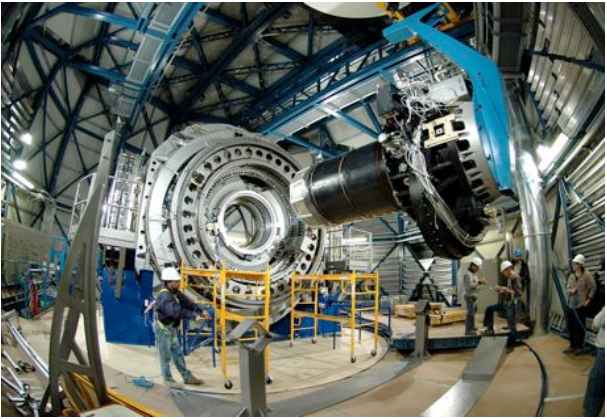 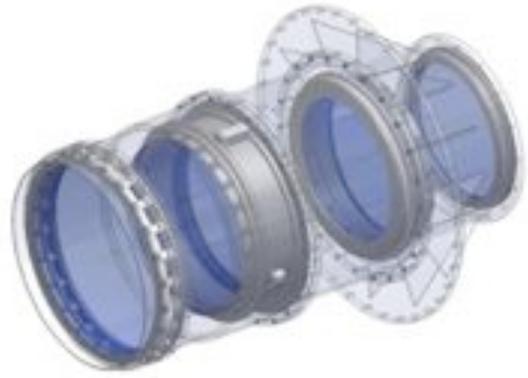

Figure 5: *Left*) Installation and removal of instruments on VISTA is performed by sliding the instrument in and out with a special lifting tool. *Right*) A VISTA WFC concept design that would slide in and out the telescope in similar fashion.

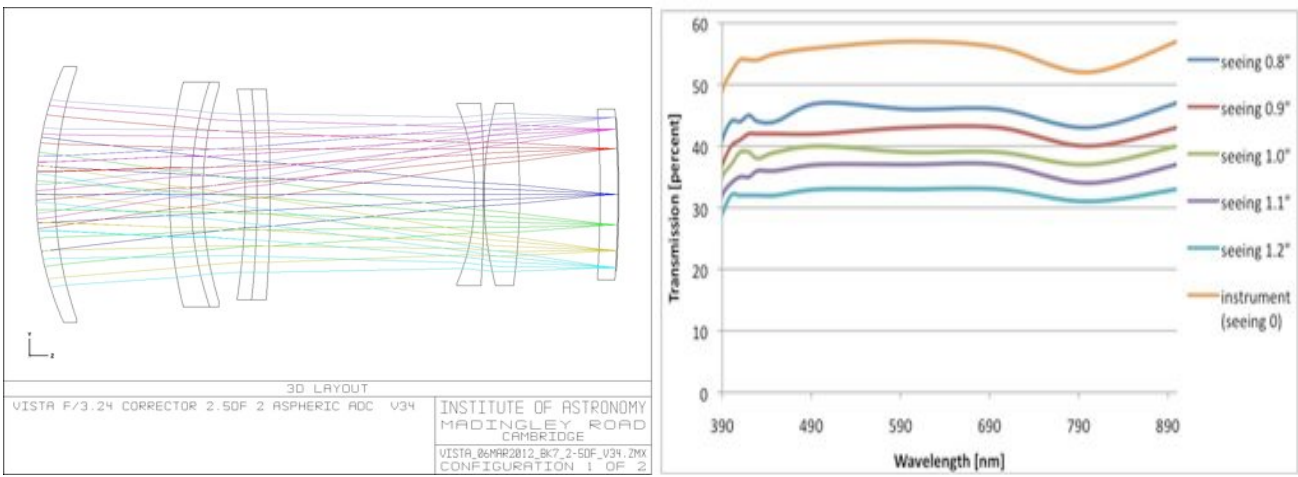

Figure 6: *Left*) Optical layout of the VISTA 2.5 degree diameter WFC design. *Right*) Calculated throughput and coupling from M1 into an 1.5 arcsec diameter uncoated fibre under listed seeing conditions analysed at 37.5° zenith distance. The oscillations in the blue are due to quantization noise of the simulations.

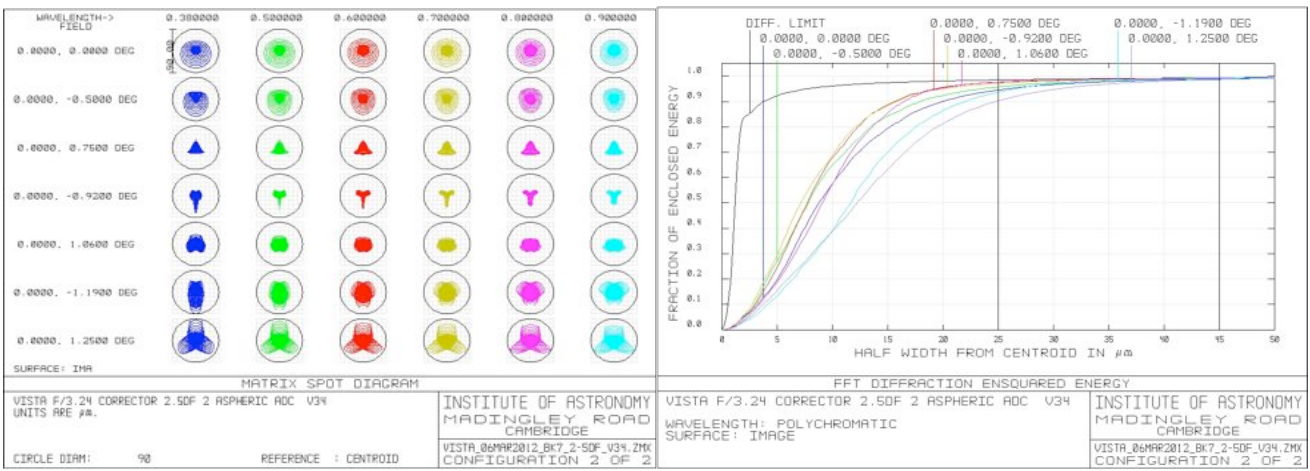

Figure 7: *Left*) Spot diagrams as function wavelength (horizontal) and distance from field centre (vertical) for the 2.5 degree diameter VISTA WFC design at zenith distance of 65°. *Right*) The ensquared energy as function of radius from image centroid at the indicated positions in the field.

## 5.2 Positioners and fibre system

Two fibre positioner concepts are under consideration for 4MOST, both based on micro fibre manipulators positioned in a fixed triangular grid pattern. The first positioner concept is a Phi-Theta system developed by the University of Munich (MuPos) and is similar to the one used on the LAMOST telescope[11]. In this system each fibre manipulator has two rotational axis, the first rotation axis giving 360° coverage in a patrol field, the second being offset driving an arm giving coverage within the patrol area (Figure 8). The development details of this positioner concept are described elsewhere in these proceedings by Lang-Bardl et al.[12].

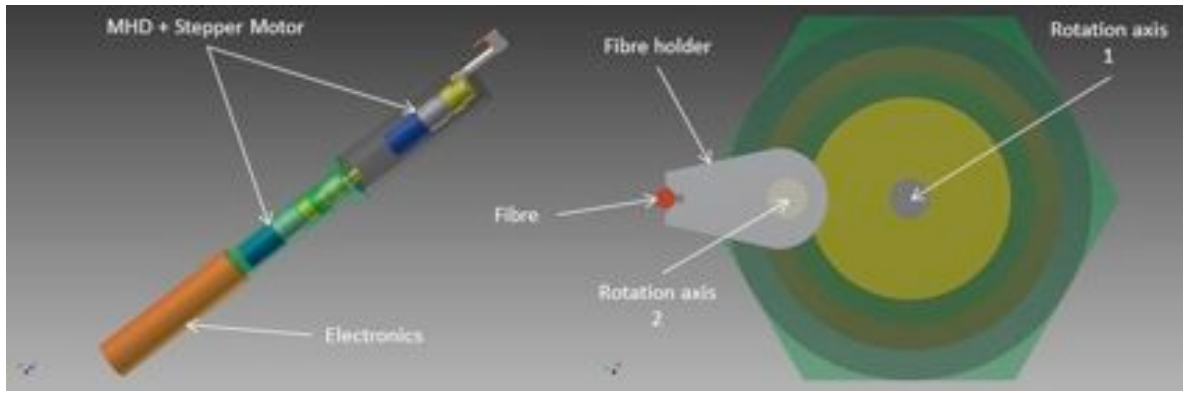

Figure 8: Mechanical design and main components of the Phi-Theta fibre manipulator of the MuPos positioner. The rotation axes of the manipulator are driven by stepper motors from Faulhaber attached to a micro harmonic drives of Micromotion.

The second positioner concept is a new R-Theta system under development at the AIP in Potsdam (PotsPos) with each fibre manipulator having a linear and a rotational movement. The PotsPos architecture was driven by Science Requirements for optimal target allocation efficiency (i.e fibre fill factor) and accommodating a fibre density for HR spectrographs of about $1/5^{th}$ to a $1/10^{th}$ compared to that of the LR spectrographs. Given the clustered nature of some science targets the ideal configuration would be a plug-plate or pick-and-place systems. However, this is extremely hard to implement with an automated, robotic system for fibre counts above about 1000, so a grid-based configuration was eventually pursued. In order to provide greater fibre-target assignment flexibility a large patrol area is assigned to each fibre actuator, such that both the LR and HR fibres have good focal surface fill factor.

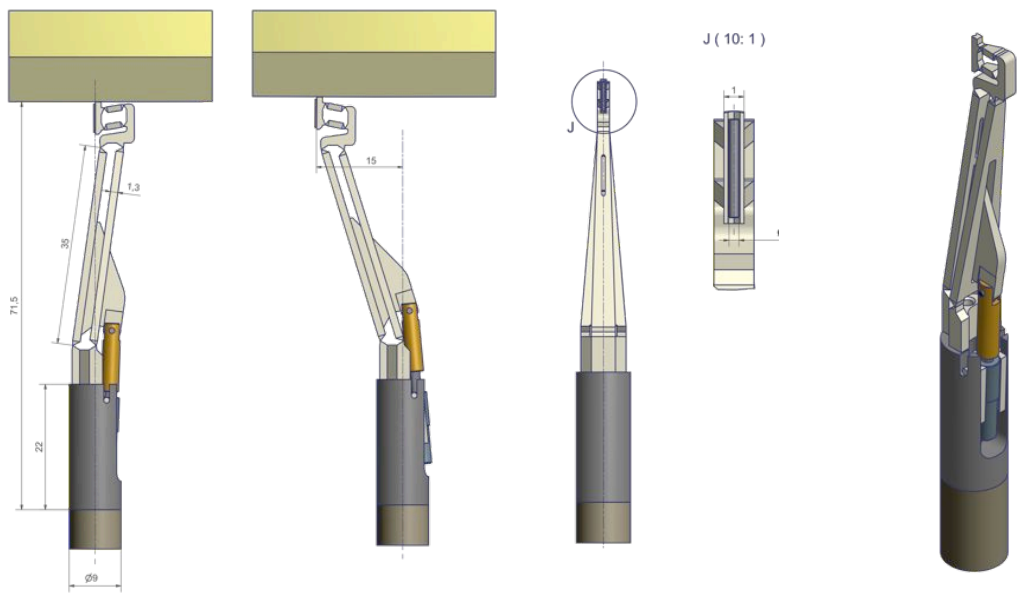

Figure 9: PotsPos actuator design, with pre-loaded flexures and Squiggle motors for both R and Theta. Linear moves (left two figures) maintain fibre focus by following the glass surface of the last lens element of the wide field corrector.

In order to maximise patrol area and minimise the collision avoidance zone the PotsPos actuators (Figure 9) have minimal cross section and unfolds in the vertical plane, perpendicular to the focal surface, instead of the horizontal employed by MuPos. This allows greater assignment flexibility, close target packing and a relatively simple collision approach. Of the 5 degrees of freedom x, y, z (focus), tip & tilt (i.e. telescope pupil pointing), only x and y need to be under active control, the others are passively controled. The configuration is an R-theta system and is in positioning capabilities similar to the moving spines system as implemented in the Echidna fibre positioning system[1], but avoids the defocus and fibre tilt light losses inherent to that design. The pre-loaded top flexure constrains the fibre pointing and uses the last element of the telescope WFC as a glass reference surface to accurately constrain the fibre focus. There is a gap between the fibre face and glass surface, which along with AR coatings, minimises the impact of Etalon fringing. The asymmetric lower parallelogram ensures the fibre stays perpendicular to the curved glass surface and allows minimal cross-section at the actuator tip for efficient allocation of adjacent targets. Three different paths to manufacture the manipulator in plastic or metal are being pursued.

The actuator patrol radii are about 1.2–1.5 grid separations and the patrol areas are such that each point in the focal surface can be reached by at least seven actuators. With just one-in-nine fibres going to the high-resolution spectrograph a large fraction of the focal surface can be reached by an HR fibre; with an one-in-six configuration the full focal surface can be reach with HR fibres. The ratio of HR to LR fibres is currently being optimised, using the facility simulator, to the maximise return for the different science programs.

Each actuator has a linear motor for R motion and a rotational motor for Theta Motion. ASIC controlled NewScale Technology piezo resonance motors where chosen, as they provide direct drive without backlash, high resolution, and high torque in a very small package. R-Motion uses the SQL-RV-1.8 linear motor, providing 0.5N (~50g) force, 0.5μm resolution, and 6mm travel in a 1.8x1.8x6mm package. Theta-motion draws on the rotational actuator developments by NewScale for the JPL Cobra fibre positioner. The motor should meet the resolution and torque requirements, but there is scope to develop a slightly larger hollow shaft motor (9mm versus 7.7mm diameter), to elegantly accommodate the fibre. The actuators will be highly modularized to facilitate manufacturing, integration, testing, and maintenance. Sets of actuators, fibres, electronics and cooling will be housed in independent modules that can be exchanged readily. Modularization of the fibre connectors is enabled by the use of commercially available US Conec MTP connectors that allow bundles of tens of fibres to be easily connected with high throughput and low focal ratio degradation and fringing when carefully manufactured and used with a gel or oil coupling[9].

The positioner Metrology Unit will determine the position of each fibre and consists of one to four cameras (mounted on the secondary mirror spider) that image the back illuminated fibre cores. A fixed grid of backlit reference fibres will be incorporated in the focal surface providing the means to determine positioner geometry/distortions and the sky coordinate geometry/distortions will be modelled and confirmed using the Acquisition & Guiding units. An IR Metrology Unit that can monitor fibre locations even *during* exposures is being considered in order to facilitate closed-loop tweaking of the fibre positions to compensate for changing atmospheric diffraction during an exposure.

## 5.3 Spectrographs

Spectrographs location
The spectrographs will be located on the telescope fork underneath the azimuth floor, rotating with the telescope (Figure 10). This allows for a short fibre run from the positioner to the spectrographs. The total number of spectrographs is still a trade-off decision depending on the number of fibres per spectrograph (affected by the neighboring spectra cross-talk constraints), the total number of fibres, and the ratio of high-to-low resolution fibres.

Low resolution spectrographs
The 4MOST low-resolution spectrograph unit developed by ASTRON/NOVA and RAL is a dual-beam system with a fibre slit input. The beam from the fibres illuminates a spherical off-axis collimator mirror that is matched to the output of the fibres and produces a 180 mm diameter beam. A dichroic beamsplitter, located after the collimator mirror, divides the beam into a blue and red arm. The beam in each arm passes through a set of collimator corrector lenses and is dispersed by a volume phase holographic grating. The resulting dispersed beam is accepted by a transmissive camera that produces a spectral image on a 6k (spatial) x 8k (spectral) CCD mosaic with 15 μm pixels. A shutter system, actuated by the detector controller, is located within each spectrograph arm, before the camera. Within the spectrograph there is a means of back-illuminating the fibre slit to provide light from each fibre at the telescope focal plane that can be detected by the fibre positioner metrology system. Each spectrograph provides means to illuminate the detector, so that detector response changes can be monitored (i.e., detector flats) independent of the spectrograph. The optical and mechanical layout is shown in Figure 11.

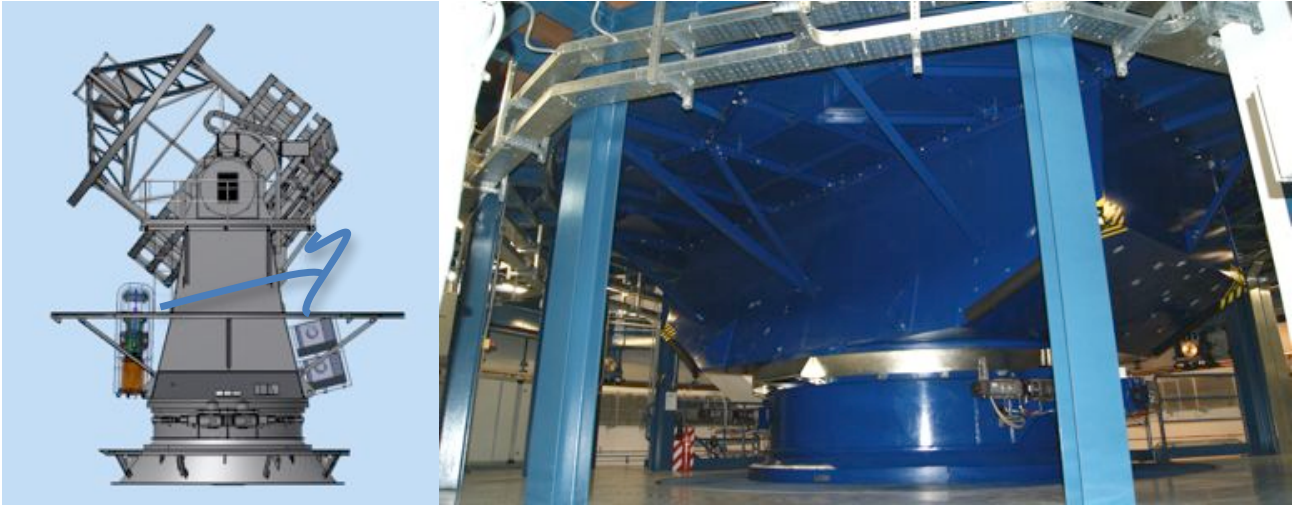

Figure 10: *Left)* Schematic location of the HR (left) and LR (right) spectrographs on the telescope fork, and the length of the fibre run (~8 and 12 m) in blue lines. *Right)* Image of area underneath the azimuth floor on one side of the telescope fork.

The design approach is to fix the nominal resolution at the centre of the blue and red wavebands and to apply constraints on the fibre input, beam diameter and detector geometry; the camera f/number is then used to control the wavelength coverage and fibre core sampling, and is set to f/1.74. This fixes the incident and diffracted angle (set equal) at 24.1°, resulting in a grating with a size of 197 mm x 180 mm and with 1627 (blue) and 1084 (red) lines/mm. The grating is feasible to manufacture with existing technologies. The sign of the grating orders is chosen so that slit curvature may be corrected in both arms simultaneously. The wavelength coverage (shown in Figure 12) is chosen to allow the 390–950 nm goal to be reached with an overlap of 20 nm for a dichroic transition at around 600 nm. The design allows up to 1000 fibres to be placed in the slit with a centre-to-centre spacing of 159 μm. Assuming a total fibre diameter of 130 μm this permits a spacing of 29 μm between the physical outer edges of the fibres.

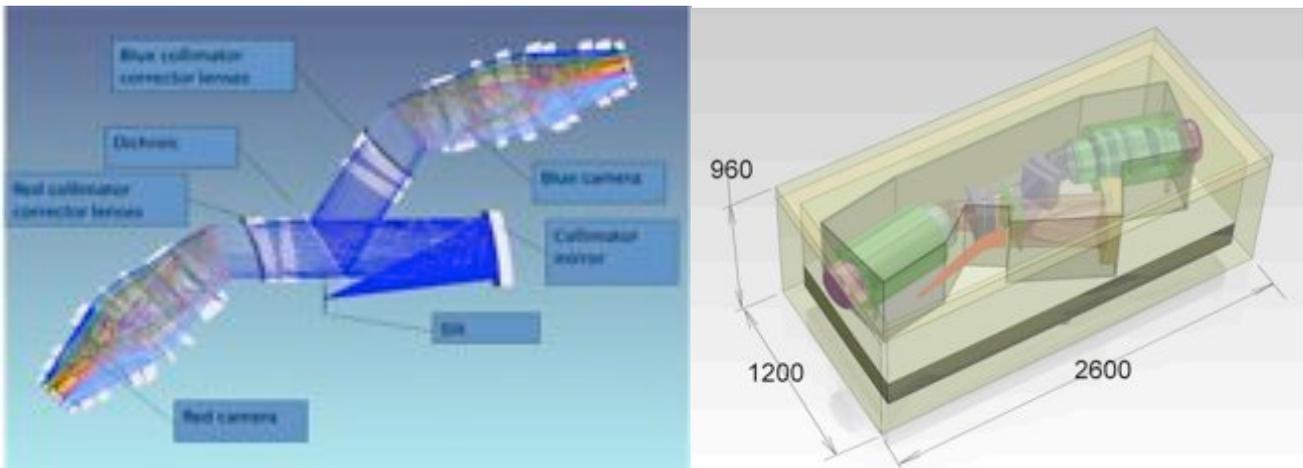

Figure 11: The optical (left) and mechanical (right) layout of the low-resolution spectrograph.

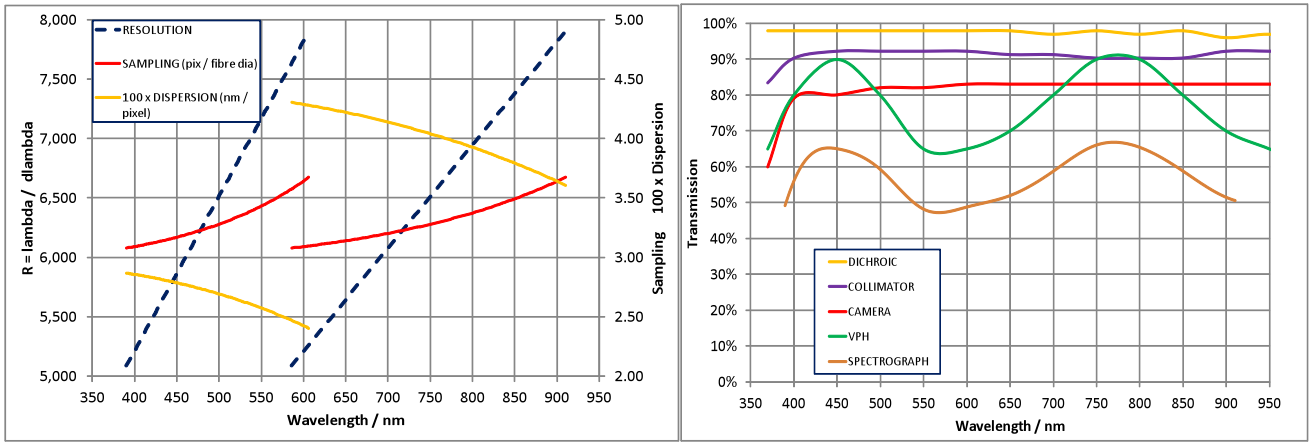

Figure 12: *Left*) Resolution, sampling and dispersion vs. wavelength. *Right)* Throughput vs. wavelength of the low-resolution spectrograph.

The spectrographs are designed for high availability and low maintenance. The amount of moving parts are minimised to only the shutters and the necessary valves for the continuous coolant flow detector cryostats. Calibration lamps and/or backlights will be of long lifetime (e.g. LED), redundant or quickly exchangeable.

High resolution spectrographs

The 4MOST high-resolution spectrograph (HRS) design as developed by GEPI, Paris, is driven by provision of a high spectral resolution over a large spectral range and by the need for coherence with the 4MOST low-resolution capability. The HRS and the LRS should compliment from an operations point of view for simultaneous observations and design-wise for: installation on telescope, common building blocks (e.g. detectors, NGC, cryostats based on those of MUSE, PLC, sensors) and shared resources (e.g. continuous flow LN2 cooling, detector and cryostat control software). With these precise objectives, as well as reliability concerns for survey use, the HRS design is a fixed configuration dual-arm spectrograph operating without active compensation of ambient effects (e.g. temperature).

A novel approach is being investigated to map the pseudo-slit loci to the sky or the calibration group with full flexibility. It relies on the use of variable optical attenuators that act as independent shutters and allow fine tuning the intensity of the light injected at the calibration ends of the fibre feed and of two-to-one combiners to route the light from these two origins to a single fixed pseudo-slit (Figure 13). Other options are being considered should this approach be impractical.

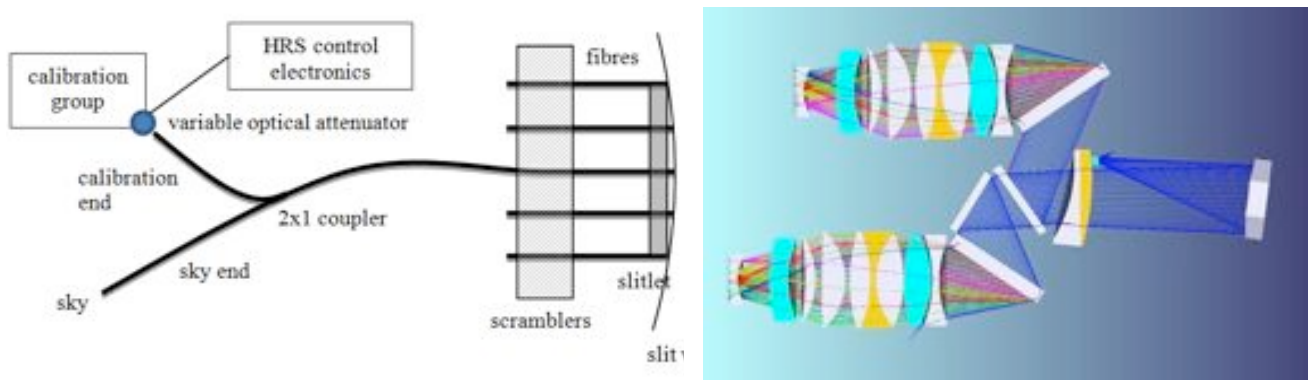

Figure 13: *Left*) Fibre feed interface coupling the light of both the sky and the calibration source into the spectrograph and slitlet organization of the pseudo-slit (slit window curvature is greatly exaggerated). *Right*) The two-arm optical layout of the HRS.

The spectrograph follows a classical two-arm design as illustrated in Figure 13. The design was mainly driven by two challenges: achieving an image quality compatible with the specified spectral resolution and avoiding large optics after the gratings. The basic features of the spectrograph concept are outline below.

- The pseudo-slit illuminates a Matsukov-type collimator that works at the same focal ratio as the beam coming out the fibres (F/3.0) so that no micro-lenses are required. Its focal length amount to 570 mm.
- The dichroic plate reflects the blue wavelengths and transmits the red ones. The cut-off is at 520 nm and needs not be steep as the gap between the two domains is wide.
- For compactness and mechanical construction ease a folding mirror is inserted on the red arm to make the optical axes of the cameras parallel.
- VPH gratings have been selected as offering the highest efficiency over a relatively large bandwidth. The two VPH gratings are placed at the average positions of the reflected and transmitted pupil images to optimise their respective sizes and feature Bragg angles of 55.6º / 56.2º with 3875 / 2637 lines/mm (blue/red).
- The blue and red cameras are identical and work at F/1.33 with a focal length of 252 mm. They feature 9 lenses of which the last is also intended to serve as the window to the cryostat (hence the distance between the penultimate and ultimate lenses).
- The image area is 116+6 mm / 111+6 mm (spectral blue/red) and 23.4 mm (spatial) with 165 fibres separated by 3.8 core diameters (axis-to-axis).

The predicted performance of the HRS is shown in Figure 14. More details about the HRS can be found elsewhere in these proceedings from Mignot et al.[13].

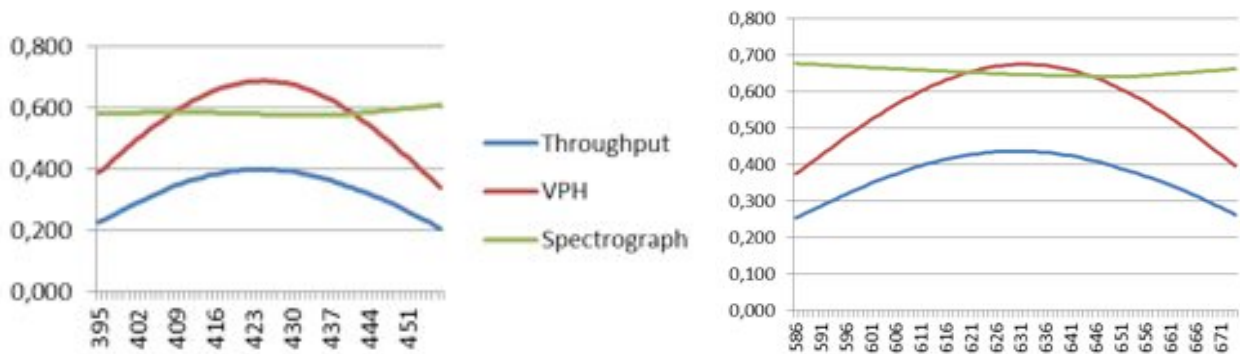

Figure 14: The predicted throughput of the blue (left) and the red (right) channel of the high resolution spectrograph.

## 6. FACILITY SIMULATOR

The 4MOST consortium has created the 4MOST Facility Simulator (4FS) in order to validate the science performance of the system and to optimize the instrument design, the science cases and the survey strategy. The simulator is expected to evolve into the planning tool needed for actual execution of the 4MOST surveys on the mountain.

Central in the 4FS is the Operations Simulator developed by MPE, Garching and described in more detail elsewhere in these proceedings by Boller et al.[3]. In its final version it will realize in a statistical manner 4MOST operations on a night by night basis, taking into account moon phase, visibility, seeing, etc. The targeting algorithm that assigns fibres to targets in a given pointing is part of the Operations Simulator. Another aspect of the instrument performance simulation is the throughput model and exposure time calculator (ETC) that calculates the detected spectrum and its signal-to-noise by modeling the atmosphere and the throughput of all instrument components for individual spectra. This modeling is performed at Observatoire de Paris à Meudon (GEPI) and is described in more detail elsewhere in these proceedings by Sartoretti et al.[15]. The third component of the 4FS is the Data Quality Controls Tool developed at the Institute of Astronomy (Cambridge), which is used to determine the success of individually simulated targets and to quantify survey progress and survey success for the different science cases.

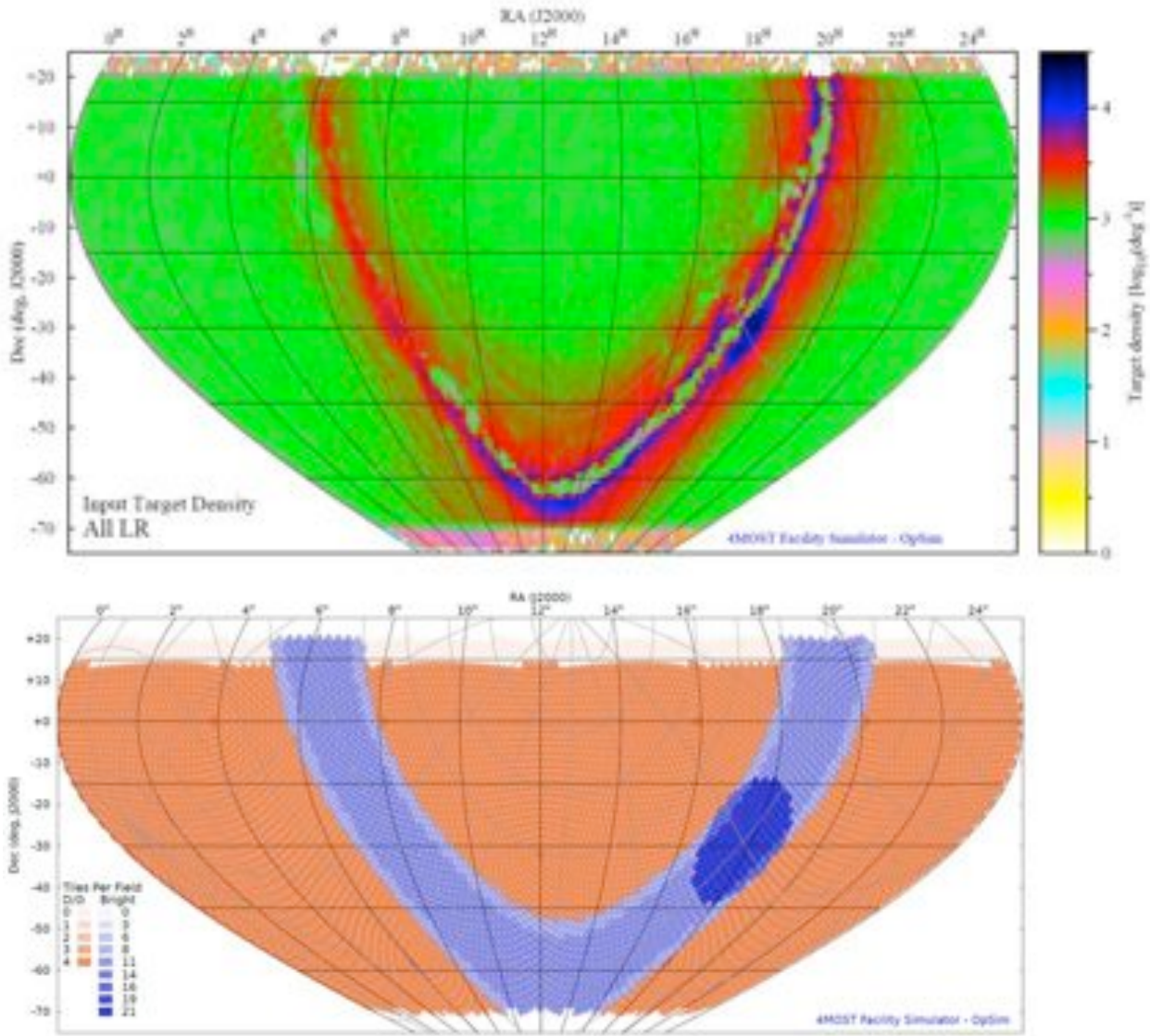

Figure 15: *Top)* The target density of all low-resolution science cases combined. The target density increases by factor 10 toward the Galactic disk, and effect that is even stronger for the high-resolution science, and argues for a very large field-of-view as this allows certain areas of the sky to be repeated several times. *Bottom)* Preliminary tiling strategy on the sky with red hexagons used for dark time and blue for bright time. The shading indicates the total number of 20 minute exposures used per pointing.

The simulator takes as input mock catalogues of seven "Design Reference Surveys (DRSs)", based on *key science projects that put the strongest constraints on the design*. The mock catalogues consist of template spectra of a limited set of typical targets of each science case plus a target distribution on the sky. Each DRS has also criteria for "observing" success of the individual spectra and a Figure of Merit for a 5-year survey. The distribution of the low-resolution targets and how the simulator tiles the sky is shown in Figure 15. An example of the targeting algorithm in the focal plane and the x,y offsets of an individual fibre are shown in Figure 16.

Initial simulations taking reasonable estimates of the overheads and calibrations into account show that with a 2500 fibre system we can observe about 20 million low-resolution and 2 million high-resolution targets in a 5-year survey when using a 1:4 ratio of high to low resolution fibres.

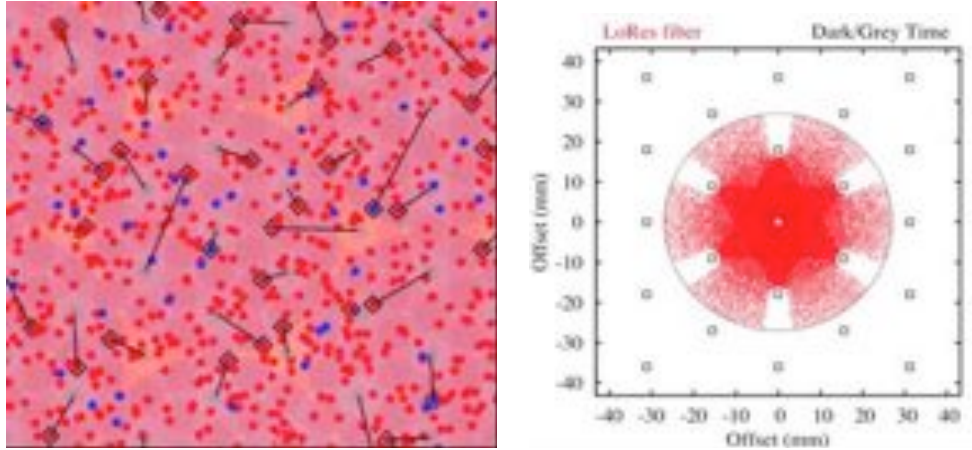

Figure 16: *Left)* The fibre-to-target assignment simulation in a crowded field region. Red targets are low-resolution spectrograph targets, blue high-resolution, lines indicate fibre assignments with selected targets squared, the background shading indicates the patrol areas of the high- and low-resulution fibres (many overlaps for this PotsPos case). *Right)* All target x,y offset locations (red points) for one particular fibre after a 5-year survey for the PotsPos positioner. This clearly shows the shadowing effect of surrounding fibre actuators, whose central positions are indicated by black circles.

## 7. SCHEDULE

The 4MOST consortium will continue the Conceptual Design study for the rest of 2012 and submit the results to ESO by Feb 1, 2013. At this point the study will be compared to the MOONS concept (described elsewhere in these proceedings by Cirasuolo et al.[4]). The decision which project(s) will go ahead is expected Spring 2013. The two concepts go on different telescopes and therefore it is technically feasible to have both projects go forward, which would open up capabilities equivalent to about 500 X-Shooters to the ESO community.

In the rest of the Conceptual Design study design and technical feasibility of critical elements of the system will be developed further and prototyped where necessary. Special attention will be paid to the positioner concepts with testing of the accuracy and reliability of the fibre actuators, characterization of the fibre throughput under stress, and the achievable accuracy of the fibre metrology system.

The continued schedule calls for an installation and commissioning of 4MOST on VISTA in late 2018 after the completion of the current VISTA IR imaging surveys. This schedule is well matched to the data releases of Gaia and eROSITA such that well-defined catalogues can be used for the 4MOST surveys.